\documentclass[12pt]{article}
\usepackage{amssymb,multicol,amscd,amsmath}
\usepackage{hyperref}

\makeatletter \@addtoreset{equation}{section} \makeatother

\def\ads{AdS_{d}}

\def\be{\begin{equation}}
\def\ee{\end{equation}}

\def\bee{\begin{eqnarray}}
\def\eee{\end{eqnarray}}

\def\bal{\begin{align}}
\def\eal{\end{align}}

\def\ba{\begin{array}}
\def\ea{\end{array}}

\def\nn{\nonumber}

\def\d{\partial}
\def\dps{\displaystyle}
\def\bpsi{\bar{\psi}}

\def\btheta{\bar{\theta}}
\def\bkappa{\bar{\kappa}}

\def\bs{\bar{s}}
\def\bv{\bar{v}}
\def\bta{\bar{\eta}}

\def\bn{ \bar{n}}
\def\bu{ \bar{u}}

\def\rvac{|0\rangle}
\def\lvac{\langle 0|}

\def\lb{\label}

\newcommand{\diff}{\mathsf{D}_0}
\newcommand{\extdiff}{{\rm d}}
\newcommand{\Eop}{\mathsf{E}_0}
\newcommand{\der}{\partial}
\newcommand{\ptl}[1]{\frac{\der}{\der#1}}
\newcommand{\pptl}[2]{\frac{\der^2}{\der#1\der#2}}

\def\cA{\mathcal{A}} \def\cB{\mathcal{B}} 
\def\cD{\mathcal{D}} \def\cE{\mathcal{E}} \def\cF{\mathcal{F}}
  
 \def\cK{\mathcal{K}} 
\def\cM{\mathcal{M}}  
\def\cP{\mathcal{P}} \def\cQ{\mathcal{Q}} \def\cR{\mathcal{R}}
\def\cS{\mathcal{S}} \def\cT{\mathcal{T}} \def\cU{\mathcal{U}}
 \def\cW{\mathcal{W}}

\newcommand{\ga}{\alpha}
\newcommand{\gb}{\beta}
   
\newcommand{\gd}{\delta}    \newcommand{\Gd}{\Delta}
\newcommand{\gep}{\epsilon}

\newcommand{\gl}{\lambda}   
\newcommand{\go}{\omega}    \newcommand{\Go}{\Omega}

\newcommand{\un}{{\underline n}}

\begin{document}


\begin{flushright}
{\small FIAN/TD/20/04}
\end{flushright}


\begin{center}

{\bf \Large Lagrangian Formulation for Free\\
\vspace{3mm} Mixed-Symmetry Bosonic Gauge Fields in $(A)dS_d$}

\vspace{1.2cm}

K.B. Alkalaev, O.V. Shaynkman and M.A. Vasiliev

\vspace{6mm}

{\it I.E.Tamm Theoretical Department, P.N.Lebedev Physical Institute, \\
Leninsky prospect 53, 119991, Moscow, Russia}

\vspace{3mm}
\texttt{alkalaev@lpi.ru}$\;\;\;$\texttt{shayn@lpi.ru}$\;\;\;$\texttt{vasiliev@lpi.ru}

\vspace{18mm}

\hspace{6.5cm}\textit{To the memory of Anatoly  Pashnev }

\vspace{18mm}

\begin{abstract}

Covariant Lagrangian formulation for free bosonic massless fields
of arbitrary mixed-symmetry type in $(A)dS_d$ space-time is
presented. The analysis is based on the frame-like formulation of
higher-spin field dynamics \cite{ASV} with higher-spin fields
described as $p$-forms taking values in appropriate modules of the
$(A)dS_d$ algebra. The problem of finding free field action is
reduced to the analysis of an appropriate differential complex,
with the derivation $\cQ$ associated with the variation of the
action. The constructed action exhibits additional gauge
symmetries in the flat limit in agreement with the general
structure of gauge symmetries for mixed-symmetry fields in
Minkowski and $(A)dS_d$ spaces.

\end{abstract}

\end{center}

\section{Introduction}

The problem of finding manifestly covariant action for free
higher-spin (HS) massless fields is to large extent the problem of
identification of their gauge symmetries. The remarkable example
is provided by Fang-Fronsdal theory of symmetric 4$d$ HS
gauge fields developed both on the flat \cite{fronsdal_flat} and
$AdS_4$ \cite{fronsdal_ads} backgrounds that was later discussed
and extended to any dimension within various
approaches \cite{Curtright:1979uz}-\cite{Buchbinder:2004gp}. For
mixed-symmetry gauge fields the problem becomes considerably more
involved since, generically, the set of gauge symmetries required to
describe an irreducible massless field is different for Minkowski and  $AdS$
spacetimes\footnote{Note that we restrict our consideration by
elementary relativistic fields corresponding to the UIR's of the
space-time symmetry algebra with the energy bounded from below.
Beyond this class, the interesting case of the partially massless
gauge fields is described in \cite{deser}.} \cite{Metsaev,BMV}. Although
there exist several formulations of the flat mixed-symmetry gauge
field dynamics \cite{curt}-\cite{MedHull} the elaboration of a
covariant formulation for generic massless fields in the $(A)dS_d$
was not known apart from numerous particular cases
\cite{ASV,BMV,SS2,siegel,Zinoviev,deMedeiros:2003px,A2}. The light-cone
formulation of generic massless fields in $\ads$ was obtained, however, by
Metsaev \cite{Metsaev,Metsaev_d5}.

By generalizing the frame-like formulation of symmetric fields developed in
\cite{LV,vf,VD5} it was proposed in Ref.
\cite{ASV} to describe a generic massless field
propagating on the $(A)dS_d$ background as a $p$-form taking values in
appropriate finite-dimensional module of the $(A)dS_d$ algebra ($o(d-1,2)$) $o(d,1)$.
This scheme provides a natural realization of all gauge symmetries and trace
conditions imposed on a mixed-symmetry gauge field in $(A)dS_d$.
In addition, the method allows one to construct  manifestly gauge invariant (linearized)
HS curvatures. In this letter we announce construction of Lagrangian formulation
for a bosonic gauge
field of any spin (\textit{i.e.},  symmetry type) in $AdS_d$ within the approach of \cite{ASV} \footnote{Although
bosonic fields in $dS_d$ space can be described analogously to the $AdS_d$ case,
for definiteness we confine ourselves to the $AdS_d$ case in this letter.}.

The paper is organized as follows. In sect. \ref{Sct.:2}
a short review of the frame-like formulation of mixed-symmetry fields is given.
In subsect. \ref{2_2} following \cite{ASV}, we introduce mixed-symmetry bosonic fields, describe
their gauge symmetries and gauge invariant linearized curvatures.
In subsect. \ref{2_3} we reformulate all these objects in the manifestly
antisymmetric basis for mixed-symmetry tensors most appropriate to the
construction of the action functional.
In sect. \ref{Sct.:3}  we introduce a convenient oscillator approach.
In sect. \ref{Sct.:4} a general form of manifestly covariant and gauge invariant
HS action generalizing the MacDowell-Mansouri action for gravity \cite{MM}
is formulated and the decoupling condition
guaranteeing that the action describes the correct number of degrees of freedom
is imposed. In sect. \ref{Sct.:5} and sect. \ref{Sct.:6} we reformulate the problem of finding
free action in terms of the analysis of a certain differential complex.
In sect. \ref{Sct.:7} we study the flat limit of the constructed action and
show that it possesses all necessary additional flat gauge symmetries absent
in the $AdS$ space \cite{Metsaev,BMV}. Conclusions are
given in sect. \ref{Sct.:8}.

\section{Frame-like formulation of mixed-symmetry fields\lb{Sct.:2}}

In the frame-like formulation of Einstein gravity with the
cosmological term the gauge field $1$-form\footnote{We work within the mostly
minus signature and use notations $\underline{m},\underline{n} =
0\div d-1\;$ for world indices, $a,b= 0\div d-1$ for tangent
Lorentz $o(d-1,1)$ vector indices and $A,B = 0 \div d $ for
tangent $\ads$ $o(d-1,2)$ vector indices. We also use condensed
notations of \cite{V1} and denote a set of symmetric vector
indices $(a_1 \ldots a_s)$ and a set of antisymmetric vector
indices $[b_1 \ldots b_p]$ as $a(s)$ and $b[p]$, respectively.}
$\Go^{AB}=-\Go^{BA}=\extdiff x^{\underline{n}}\Go^{AB}_{\underline{n}}$ is
associated with basis elements of the $AdS_d$
algebra.
With the help of the compensator field $V^A(x)$ normalized as
$V^A V_A = 1$ \cite{compensator}, $\Go^{AB}$ can be
covariantly decomposed into the frame field $\lambda E^A = \extdiff V^A+\Go^{AB}V_B$ and the spin
connection $\omega^{AB} = \Go^{AB} -\lambda\,(E^{A}\,V^{B}-E^{B}\,V^{A})$. Parameter $\lambda$
is the inverse radius of the $AdS_d$ spacetime. (Note that $\lambda$ has to be introduced
to make the frame field $E^A$ dimensionless.) The following relations hold
$E^A\,V_A = 0$, $\cD V^A \equiv \extdiff V^A+\omega^{AB}V_B= 0$.   In these notation, the background $\ads$ geometry  is
described by the gauge field
$\Go_0^{AB}=(E_0^A$, $\omega_0^{AB})$,
which satisfies  the non-degeneracy condition $rank(E_{0\un}^A)=d$ and
the zero-curvature equation
\be
\label{zerocurv}
R^{AB}(\Go_0)\equiv \extdiff \Go_0^{AB}+\Go_0^A{}_C\wedge \Go_0{}^{CB}=0\,\Longleftrightarrow\,
D_0^2=0\;,
\ee
where $D_0 T^{AB\ldots} \equiv \extdiff T^{AB\ldots}+ \Go_0{}^{A}{}_C\wedge T^{CB\ldots} +
\Go_0{}^{B}{}_C\wedge T^{AC\ldots} + \cdots$ (see, e.g., \cite{ASV,VD5} for more details).
The $\ads$ metric tensor is expressed in terms of the frame field $E^A_0$ as
$g_{\underline{mn}}=E_{0\underline{m}}{}^A E_{0\underline{n}}{}^B\,\eta_{AB}\,.$

\subsection{Mixed-symmetry bosonic frame-type gauge fields\lb{2_2}}

Let us consider a massless bosonic field $\Phi(x)$ corresponding to a spin ${\bf s}$
unitary module of Poincare ($iso(d-1,1)$) or $AdS_d$ ($o(d-1,2)$)
algebra in $d$ dimensions.
A spin ${\bf s}$ of a bosonic field is described by a set of integers
$
{\bf s}=(\underbrace{s,\ldots,s}_p,s_{p+1},\ldots,s_q)
$,
where $s>s_{p+1}\geq\ldots\geq s_q$, $1\leq p\leq q$ and $q=[(d-2)/2]$ for Minkowski case, or
$q=[(d-1)/2]$ for the $AdS_d$ case.
$\Phi(x)$ can be described as a tensor field
$
\Phi^{a_1(s),\ldots,a_q(s_q)}(x)
$,
which satisfies certain tracelessness conditions and
Young symmetry conditions corresponding to the Lorentz Young tableau
$Y(s,\ldots,s_q)$ with $q$ rows of the lengths $s,\ldots,s_q$,
{\it i.e.} it is symmetric in each group of indices $a_i(s_i)$
and its symmetrization over all indices from a given group $a_i(s_i)$ with an index $a_j$
gives zero once $j>i$.

In Minkowski space, the physical degrees of freedom described by the field $\Phi(x)$ are characterized by a
unitary finite-dimensional irrep $V({\bf s})$ of the Wigner little algebra $o(d-2)$.
Redundant degrees of freedom are removed by appropriate
gauge symmetries \cite{LabastidaMorris,Labastida2}
\be\label{flat_g}
\gd\Phi^{a_1(s),\ldots,a_i(s_i),\ldots,a_q(s_q)}=\sum_{i=p}^q\cP_\Phi(\d^{a_i}
S_i^{a_1(s),\ldots,a_i(s_i-1),\ldots,a_q(s_q)})\,,
\ee
where gauge parameters $S_i^{a_1(s),\ldots,a_i(s_i-1),\ldots,a_q(s_q)}$ correspond
to all allowed (\textit{i.e.} non-zero) Young tableaux $Y(s,\ldots,s_i-1,\ldots,s_q)$
and $\cP_\Phi$ is the projector on $Y(s,\ldots,s_q)$.

In the $AdS_d$ space, the physical degrees of freedom of the field $\Phi(x)$
are characterized by a unitary finite-dimensional irrep $V(E_0,{\bf s})$ of the maximal
compact subalgebra $o(2)\oplus o(d-1) \subset o(d-1,2)$, where $E_0$ is the lowest energy eigenvalue.
By definition, the lowest energy of a massless field $E_0 =s-p+d-2$ is such
that the $\ads$ representation $D(E_0,{\bf s})$ induced from
$V(E_0,{\bf s})$ corresponds to the boundary
of the unitarity region in the weight space \cite{Metsaev}.
As shown in \cite{Metsaev,BMV}, in $AdS_d$ the only gauge symmetry of the field $\Phi(x)$ is associated with
$S_p$ (\textit{i.e.} with the first gauge parameter in (\ref{flat_g}))
\be
\label{ads_g}
\gd\Phi^{a_1(s),\ldots,a_p(s),\ldots,a_q(s_q)}=\cP_\Phi(\cD^{a_p}
S_p^{a_1(s),\ldots,a_p(s-1),\ldots,a_q(s_q)})\,,
\ee
while all other gauge symmetries are lost.

According to \cite{ASV} the field $\Phi(x)$ can be equivalently described
within the framework of the frame-like formalism
analogous to the MacDowell-Mansouri approach to gravity \cite{MM}, which was generalized
to totally symmetric spin $s$ gauge fields in \cite{LV,vf,VD5}.
In \cite{ASV} it was proposed that dynamics of the metric-type field $\Phi(x)$ can be described
by a $p$-form gauge field
\be \label{ads_field}
\Omega_{(p)}{}^{A_0(s-1),\,\ldots\,,A_p(s-1),\,A_{p+1}(s_{p+1}),\,
\ldots\,,A_{q}(s_q)}\,,
\ee
which takes values in the irreducible $o(d-1,2)$-module corresponding to the
Young tableau $Y(\underbrace{s-1,\ldots,s-1}_{p+1},s_{p+1},\ldots,s_q)$ constructed
by cutting the shortest column of the height $p$ of the $o(d-1,1)$ Young tableau corresponding to the
$\Phi(x)$ and then by adding the longest row of the length $s-1$.
The linearized HS curvature $(p+1)$-form associated with the gauge $p$-form
field (\ref{ads_field}) is
\be\label{adscurv}
R_{(p+1)}{}^{A_0(s-1),\, \ldots\, ,\,A_{q}(s_q)} = D_0
\Omega_{(p)}{}^{A_0(s-1),\, \ldots\, ,\,A_{q}(s_q)}\,.
\ee
The curvature $(p+1)$-form is manifestly invariant under the gauge
transformations
\be\label{adsgaugetr}
\delta\Omega_{(p)}{}^{A_0(s-1),\, \ldots\, ,\,A_{q}(s_q)} = D_0
\xi_{(p-1)}{}^{A_0(s-1),\, \ldots\, ,\,A_{q}(s_q)}\;
\ee
with the $(p-1)$-form gauge parameter
$\xi_{(p-1)}{}^{A_0(s-1),\,\ldots\,,\,A_{q}(s_q)}$
and satisfies the Bianchi identities
\be\label{Bianchi}
D_0 R_{(p+1)}{}^{A_0(s-1),\, \ldots\, ,\,A_{q}(s_q)} = 0\;
\ee
as a consequence of the zero-curvature equation (\ref{zerocurv}).

Reduction of the $p$-form gauge field (\ref{ads_field}) with respect to the Lorentz subalgebra
$o(d-1,1)\subset o(d-1,2)$ gives a set of $p$-form
Lorentz-covariant gauge fields containing physical field, auxiliary fields
and extra fields associated with different irreducible $o(d-1,1)$-modules
explicitly described in \cite{ASV}.
In particular, the physical field, being analogous to the frame field in gravity, is
\be
\label{phys_field}
\gl^{s-1}e_{(p)}{}^{A_1(s-1),\ldots, A_{q}(s_{q})}
=V_{A_0} \cdots V_{A_0}
\Omega_{(p)}{}^{A_0(s-1),A_1(s-1),\ldots,A_{q}(s_q)}\,,
\ee
\textit{i.e.} the physical field $e_{(p)}$ is the maximally (\textit{i.e.},
$s-1$ times: contraction of any $s$ indices with $V_A$ gives zero
by the  Young properties of $\Omega_{(p)}{}^{A_0(s-1),\,\ldots\,,A_{q}(s_q)}$)
$V$-tangential component of $\Go_{(p)}$. There is also a
number of $s-2$ times $V$-tangential components of $\Go_{(p)}$,
which are called auxiliary fields. All other Lorentz-covariant components of field
$\Go_{(p)}$ are called extra fields.

Using the standard gauge for the compensator field $V^A=\delta^A{}_{d+1}$
\cite{compensator} one can rewrite
(\ref{phys_field}) with Lorentz indices
$
\gl^{s-1}e_{(p)}{}^{a_1(s-1),\ldots, a_{q}(s_{q})}=V_{A_0} \cdots V_{A_0}
\Omega_{(p)}{}^{A_0(s-1),a_1(s-1),\ldots,a_{q}(s_q)}
$. The frame-type field (\ref{phys_field}) is related to the metric-type field by the formula\\
$
\Phi^{a_1(s_1),\,a_2(s_2),\,\ldots\,,\,a_{q}(s_{q})}(x)=
e_{(p)}^{a_1 \ldots a_p;\;a_1(s-1),\,\,\ldots\,, a_p(s-1),\,
a_{p+1}(s_{p+1}),\,\ldots\,, a_{q}(s_{q})}(x)
$,
where symmetrization over the indices denoted by the same
letter is assumed, \textit{i.e.} the metric-type field $\Phi(x)$ results
from symmetrization of the $p$-form indices, converted into tangent ones,
with tangent indices of the first $p$ rows of the $p$-form field $e_{(p)}$.
The gauge transformations (\ref{ads_g}) and (\ref{adsgaugetr}) are related
by the analogous formula.

It was proposed in \cite{ASV} to look for the action of a free mixed-symmetry
bosonic field in the form
\be
\label{action0}
\cS_2 = \int_{\cM^d}\; U^{\ldots}(V)\gep^{\ldots}{}_{M_1\ldots M_{d-2p-2}N}
E_0^{M_1}\wedge\cdots
\wedge E_0^{M_{d-2p-2}}V^N\wedge\; R_{(p+1)}^{\ldots} \wedge
\,R_{(p+1)}^{\ldots}\,,
\ee
where $U^{\ldots} (V)$ are some coefficients which
parameterize various types of index contractions between
curvatures, compensators and the $\gep$-tensor.
Any such action is manifestly $AdS$ covariant and gauge invariant
with respect to the gauge transformations (\ref{adsgaugetr}). In \cite{ASV,A2} several
examples of mixed-symmetry fields were described in terms of the actions
of the form (\ref{action0}). In this letter we extend these results to massless
bosonic fields of generic mixed-symmetry type.

\subsection{Antisymmetric basis\lb{2_3}}

Let us rewrite the gauge field $\Go_{(p)}$ (\ref{ads_field}) in the antisymmetric basis as
\begin{align}
&\Omega_{(p)}^{A_1[\tilde{h}_1],\,...\,,A_{s-1}[\tilde{h}_{s-1}]}\,,\label{ads_field2}
\end{align}
where $\tilde{h}_1\geq\cdots\geq\tilde{h}_{s-1}\geq p+1$ are the
heights of the columns of the $o(d-1,2)$ Young tableau corresponding
to (\ref{ads_field}). Here $A_i[\tilde{h}_i]$ denotes antisymmetrized indices
$A_i^1 .... A_i^{\tilde{h}_i}$
associated with the $\tilde{h}_i$-th column. In the antisymmetric basis it is required that antisymmetrization
over all indices from a given group $A_i[\tilde{h}_i]$ with an index $A_j$ gives zero once $j>i$.
The symmetric and antisymmetric bases are equivalent
being related by a linear map. Note that $\tilde{h}_1\leq q=[(d-1)/2]$.  It is
convenient to combine columns of equal heights into vertical
blocks. The result can be depicted as

\be\label{YT}
\bigskip
\begin{picture}(250,80)(100,128)

{\linethickness{.500mm}
\put(100,210){\line(1,0){250}}\put(100,130){\line(0,1){80}}}

{\linethickness{.500mm} \put(150,130){\line(0,1){80}}
\put(100,130){\line(1,0){50}} \put(110,115){\footnotesize $(m_1,h_1)$}}

\put(110,130){\line(0,1){80}} \put(120,130){\line(0,1){80}}
\put(130,130){\line(0,1){80}} \put(140,130){\line(0,1){80}}

\put(100,140){\line(1,0){50}} \put(100,150){\line(1,0){50}}
\put(100,160){\line(1,0){50}} \put(100,170){\line(1,0){50}}
\put(100,180){\line(1,0){50}} \put(100,190){\line(1,0){50}}
\put(100,200){\line(1,0){50}}

{\linethickness{.500mm} \put(200,150){\line(0,1){60}}
\put(150,150){\line(1,0){50}}\put(160,135){\footnotesize $(m_2,h_2)$ }}

\put(160,150){\line(0,1){60}} \put(170,150){\line(0,1){60}}
\put(180,150){\line(0,1){60}} \put(190,150){\line(0,1){60}}

\put(150,160){\line(1,0){50}} \put(150,170){\line(1,0){50}}
\put(150,180){\line(1,0){50}} \put(150,190){\line(1,0){50}}
\put(150,200){\line(1,0){50}}

\put(212,200){\circle*{2}} \put(212,190){\circle*{2}}
\put(212,180){\circle*{2}} \put(212,170){\circle*{2}}

\put(225,200){\circle*{2}} \put(225,190){\circle*{2}}
\put(225,180){\circle*{2}}

\put(238,200){\circle*{2}} \put(238,190){\circle*{2}}

{\linethickness{.500mm} \put(250,170){\line(0,1){40}}
\put(250,170){\line(1,0){50}} \put(300,170){\line(0,1){40}}
\put(247,155){\footnotesize $(m_{n-1},h_{n-1})$}}

\put(260,170){\line(0,1){40}} \put(270,170){\line(0,1){40}}
\put(280,170){\line(0,1){40}} \put(290,170){\line(0,1){40}}

\put(250,180){\line(1,0){50}} \put(250,190){\line(1,0){50}}
\put(250,200){\line(1,0){50}} \put(250,210){\line(1,0){50}}

{\linethickness{.500mm} \put(350,190){\line(0,1){20}}
\put(300,190){\line(1,0){50}} \put(310,175){\footnotesize $(m_n,h_n)$}}

\put(310,190){\line(0,1){20}} \put(320,190){\line(0,1){20}}
\put(330,190){\line(0,1){20}} \put(340,190){\line(0,1){20}}

\put(300,200){\line(1,0){50}}

\end{picture}
\ee
Here  $I$-th block has size $(m_I,h_I)$, where $m_I$ is its length and $h_I$ is its
height, $I=1\div n$. We have
$
q\geq h_1=\tilde{h}_1=\ldots= \tilde{h}_{m_1} >
h_2=\tilde{h}_{m_1+1} = \ldots =
\tilde{h}_{m_1+m_2}>\ldots>
h_n = \tilde{h}_{m_1+ \ldots +m_{n-1}+1} = \ldots
= \tilde{h}_{s-1}\geq p+1
$.

Reduction of the traceless $o(d-1,2)$ Young tableau
(\ref{YT}) with respect to the Lorentz subalgebra
$o(d-1,1)\subset o(d-1,2)$ gives the set of traceless
$o(d-1,1)$ Young tableaux of the following form
\be\label{rYT}
\bigskip
\begin{picture}(250,85)(100,127)

{\linethickness{.500mm}
\put(100,210){\line(1,0){250}} \put(100,130){\line(0,1){80}}}

{\linethickness{.500mm}
\put(150,140){\line(0,1){70}}
\put(100,130){\line(1,0){30}}
\put(130,130){\line(0,1){10}}
\put(130,140){\line(1,0){20}}
\put(110,115){\footnotesize $(m_1,h_1)$}}
\put(138,130){\footnotesize $t_1$}

\put(110,130){\line(0,1){80}} \put(120,130){\line(0,1){80}}
\put(130,140){\line(0,1){70}} \put(140,140){\line(0,1){70}}

\put(100,140){\line(1,0){50}} \put(100,150){\line(1,0){50}}
\put(100,160){\line(1,0){50}} \put(100,170){\line(1,0){50}}
\put(100,180){\line(1,0){50}} \put(100,190){\line(1,0){50}}
\put(100,200){\line(1,0){50}}

{\linethickness{.500mm}
\put(200,160){\line(0,1){50}}
\put(150,150){\line(1,0){20}}
\put(170,150){\line(0,1){10}}
\put(170,160){\line(1,0){30}}
\put(160,135){\footnotesize $(m_2,h_2)$}}
\put(183,150){\footnotesize $t_2$}

\put(160,150){\line(0,1){60}} \put(170,150){\line(0,1){60}}
\put(180,160){\line(0,1){50}} \put(190,160){\line(0,1){50}}

\put(150,160){\line(1,0){20}} \put(150,170){\line(1,0){50}}
\put(150,180){\line(1,0){50}} \put(150,190){\line(1,0){50}}
\put(150,200){\line(1,0){50}}

\put(212,200){\circle*{2}} \put(212,190){\circle*{2}}
\put(212,180){\circle*{2}} \put(212,170){\circle*{2}}

\put(225,200){\circle*{2}} \put(225,190){\circle*{2}}
\put(225,180){\circle*{2}}

\put(238,200){\circle*{2}} \put(238,190){\circle*{2}}

{\linethickness{.500mm}
\put(250,170){\line(0,1){40}}
\put(250,170){\line(1,0){30}}
\put(300,180){\line(0,1){30}}
\put(280,170){\line(0,1){10}}
\put(280,180){\line(1,0){20}}

\put(247,155){\footnotesize $(m_{n-1},h_{n-1})$}}
\put(283,170){\footnotesize $t_{n-1}$}

\put(260,170){\line(0,1){40}} \put(270,170){\line(0,1){40}}
\put(280,170){\line(0,1){40}} \put(290,180){\line(0,1){30}}

\put(250,180){\line(1,0){30}} \put(250,190){\line(1,0){50}}
\put(250,200){\line(1,0){50}} \put(250,210){\line(1,0){50}}

{\linethickness{.500mm}
\put(350,200){\line(0,1){10}}
\put(300,190){\line(1,0){40}}
\put(340,190){\line(0,1){10}}
\put(340,200){\line(1,0){10}}
\put(310,175){\footnotesize $(m_n,h_n)$}}
\put(343,190){\footnotesize $t_n$}

\put(310,190){\line(0,1){20}} \put(320,190){\line(0,1){20}}
\put(330,190){\line(0,1){20}} \put(340,200){\line(0,1){10}}

\put(300,200){\line(1,0){40}}

\end{picture}
\ee
with various  $t_I$ such that $0\leq t_I\leq m_I$, $I=1\div n$.

The set of Lorentz $p$-form gauge fields resulting from the
$\ads$ $p$-form field $\Go_{(p)}$ consists of
\begin{itemize}
\item Physical field $e_{(p)}$  with $t_I=m_I$, $I=1\div n$ which corresponds to the Lorentz Young tableau (\ref{rYT}) with the
minimal number of cells.

\item Auxiliary fields  with
$t_J=m_J-1$ for some fixed $J$ and $t_I=m_I$, $I\neq J$ which have
one more cell compared to the Young tableau of the physical field.
We distinguish between ``relevant'' auxiliary field $\go_{(p)}$ with $t_1=m_1-1$
and ``irrelevant'' auxiliary fields $\go'_{(p)}$ with $t_J=m_J-1$,
$J\neq 1$.

\item Extra fields $w_{(p)}$ with $\sum\limits_{I=1}^n t_I\leq s-3$ which have two or more additional cells
compared to the Young tableau of the physical field.
\end{itemize}

 The characteristic feature of the relevant auxiliary field $\go_{(p)}$ is that it
is the most antisymmetric among all auxiliary fields, {\it i.e.} its first column  is of
maximal  possible height $h_1$. This terminology for auxiliary fields is introduced
to emphasize that they play different dynamical roles as will be explained in sect. \ref{Sct.:4}.

In the antisymmetric basis, the formula (\ref{phys_field}) for the  physical field $e_{(p)}$
is replaced by
\be\label{phys}
\gl^{s-1}e_{(p)}{}^{A_1[\tilde{h}_1-1],\,\ldots\,,A_{s-1}[\tilde{h}_{s-1}-1]}
=V_{A_1} \ldots V_{A_{s-1}}
\Omega_{(p)}{}^{A_1[\tilde{h}_1],\,\ldots\,,A_{s-1}[\tilde{h}_{s-1}]}\,.
\ee
The expression for the relevant auxiliary field  is
\be
\ba{c}
\label{true_auxiliary}
\dps
\gl^{s-2}\go_{(p)}^{A_1[\tilde{h}_1],A_2[\tilde{h}_2-1],\,\ldots\,,
A_{s-1}[\tilde{h}_{s-1}-1]}=
V_{A_2}\cdots V_{A_{s-1}}
\Omega_{(p)}{}^{A_1[\tilde{h}_1],A_2[\tilde{h}_2],\,\ldots\,,A_{s-1}[\tilde{h}_{s-1}]}
\\
\\
- \tilde{h}_1V^{A_1}V_{A_2} \ldots V_{A_{s-1}}V_B \,\Omega_{(p)}{}^{BA_1[\tilde{h}_1-1],\,\ldots\,,A_{s-1}[\tilde{h}_{s-1}]}\;.
\ea
\ee
These formulae for the physical and the relevant auxiliary fields contain no explicit Young symmetry
projectors that makes the antisymmetric basis of the Young tableaux most convenient for our purposes.
Note that in the case of gravity, the  formula (\ref{true_auxiliary}) reduces to  $\omega^{AB} = \Go^{AB} -\lambda\,(E^{A}\,V^{B}-E^{B}\,V^{A})$ (see the beginning of sect. 2).

\section{Generating Fock oscillator approach\lb{Sct.:3}}

Let us introduce the set of fermionic oscillators $\psi_\alpha{}^A =
(\psi_i{}^A, \psi^{j\,A})$ and $\bpsi_\alpha{}^A = (\bpsi_i{}^A,
\bpsi^{j\,A})$ ($i,j =1\div (s-1)$, $\ga=1\div 2(s-1)$), which satisfy
the anticommutation relations
\begin{align}\label{psi_anticommutator}
&\{\psi_i{}^A, \bpsi^{jB}\}=\delta_i^j\;\eta^{AB}\,,
&&\{\psi^{iA}, \bpsi_j{}^B\}=\delta^i_j\;\eta^{AB}
\end{align}
with all other anticommutators equal to zero.
Also introduce fermionic oscillators $\theta^A$ and $\btheta^B$
which satisfy  anticommutation relations
\be
\label{theta}
\{\theta^A,\btheta^B\}=\eta^{AB}\,, \quad \{\theta^A,
\theta^B\} = 0\,, \quad \{\btheta^A, \btheta^B\} = 0
\ee
and anticommute with $\psi_\ga{}^A$ and $\bpsi_\ga{}^A$.

Let us define the left and right Fock vacua by
$\bpsi_\ga^A\rvac=0$, $\btheta^A\rvac=0$ and $\lvac\psi_\ga^A=0$, $\lvac\btheta^A=0$
along with
\begin{align}
\label{LC_sym}
\lvac \,\theta^{A_1}\cdots\theta^{A_{d+1}}\rvac=\gep^{A_1\cdots A_{d+1}}\,,
&& \lvac \,\theta^{A_1}\cdots\theta^{A_{k}}\rvac=0\,,\;\;\textrm{for}\;\;  k\neq d+1\,.
\end{align}
The oscillators $\theta$ provide a convenient way to introduce $o(d-1,2)$
$\gep$-tensor via formula (\ref{LC_sym}).

In our construction, $p$-form $o(d-1,2)$ gauge fields
will be described as vectors of two types
$|\hat{\Omega}_{(p)}\rangle=\hat{\Go}_{(p)}\rvac$ and
$|\breve{\Omega}_{(p)}\rangle=\breve{\Go}_{(p)}\rvac$, where
\begin{align}
\label{operators}
\hat{\Omega}_{(p)}&=\Omega_{(p)}{}^{A_1[\tilde{h}_1],...,A_{s-1}[\tilde{h}_{s-1}]}
(\psi^1_{A_1})^{\tilde{h}_1}\cdots(\psi^{s-1}_{A_{s-1}})^{\tilde{h}_{s-1}}\,,\nn\\
\breve{\Omega}_{(p)}&=\Omega_{(p)}{}_{A_1[\tilde{h}_1],...,A_{s-1}[\tilde{h}_{s-1}]}
(\psi_1^{A_1})^{\tilde{h}_1}\cdots (\psi_{s-1}^{A_{s-1}})^{\tilde{h}_{s-1}}\,.
\end{align}
More generally,  operators $\hat{A}_{(m)}$ and $\breve{A}_{(m)}$ will be assumed to be
analogously constructed from a $m$-form $A_{(m)}$ instead of $\Omega_{(p)}$.
The Young symmetry and tracelessness conditions on the $p$-form gauge fields can be written as
\begin{align}
\label{irr1}
l^i{}_j|\hat{\Omega}_{(p)}\rangle&=0\,,\;\;i<j\,,&\bs_{ij}|\hat{\Omega}_{(p)}\rangle&=0\,,\\
\label{irr2}
l_i{}^j|\breve{\Omega}_{(p)}\rangle&=0\,,\;\;i<j\,,&\bs^{ij}|\breve{\Omega}_{(p)}\rangle&=0\,,\\
\label{irr3}
l^i{}_i |\hat{\Omega}_{(p)}\rangle& = \tilde{h}_i|\hat{\Omega}_{(p)}\rangle,\;\;\; & l_i{}^i|\breve{\Omega}_{(p)}\rangle& =
\tilde{h}_i|\breve{\Omega}_{(p)}\rangle
\end{align}
with
\be
l_{\alpha\beta}=\eta_{AB}\,\psi_\alpha^A\bpsi_\beta^B\,,\quad
\bar{s}_{\alpha\beta}=\eta_{AB}\,\bpsi_\alpha^A\bpsi_\beta^B\,.
\ee
The linearized curvatures (\ref{adscurv}) are
\be
\label{curv}
|\hat{R}_{(p+1)}\rangle = \hat{R}_{(p+1)}\rvac=\diff|\hat{\Omega}_{(p)}\rangle\,,
\qquad |\breve{R}_{(p+1)}\rangle =\breve{R}_{(p+1)}\rvac=
\diff|\breve{\Omega}_{(p)}\rangle\,.
\ee
Here the $o(d-1,2)$ covariant background derivative is given by
\be
\diff=\extdiff+\Go_0{}^A{}_B\psi^i_A\bpsi_i^B+\Go_0{}_A{}^B\psi_i^A\bpsi^i_B+
\Go_0{}^A{}_B\theta_A\btheta^B\,,\qquad \diff^2=0\,,
\ee
where $\Go_0{}^{AB}$ is the background $\ads$ gauge field satisfying  the zero-curvature condition (\ref{zerocurv}).
The gauge transformations (\ref{adsgaugetr}) and Bianchi identities (\ref{Bianchi}) are
\begin{align}
\dps \delta|\hat{\Omega}_{(p)}\rangle &= \diff|\hat{\xi}_{(p-1)}\rangle\,, &
\diff |\hat{R}_{(p+1)}\rangle &= 0\,,\nn\\
\label{gauge+bianchi}\dps \delta|\breve{\Omega}_{(p)}\rangle &= \diff|\breve{\xi}_{(p-1)}\rangle\,,
&\diff |\breve{R}_{(p+1)}\rangle & = 0\,.
\end{align}

In the sequel we make use of the following operators
\begin{align}
\bar{\eta}_\ga=\bpsi_\ga^A\theta_A\,,&&\bar{v}_\ga=\bpsi_\ga^AV_A\,,&&
\chi=\theta^AV_A\,,&&\Eop=E_0^A\theta_A\,,
\end{align}
where $V^A$ and $E_0^A$ are the compensator and the background frame field, respectively.

\section{The higher-spin action \label{Sct.:4}}

In the antisymmetric basis, the action functional $\cS_2$ still has the form (\ref{action0}).
With the help of the Fock notations of
the previous section, it reads as
\be\label{action}
\cS_2=\int_{\cM^d}\lvac(\wedge\Eop)^{d-2p-2}\chi\,\cU(\bar{s},\bar{\eta},\bar{v})
\wedge\hat{R}_{(p+1)}\wedge \breve{R}_{(p+1)}\rvac\,,
\ee
where $\cU(\bar{s},\bar{\eta},\bar{v})$ is some polynomial of
$\bar{s}_{\ga\gb}$, $\bar{\eta}_{\ga}$ and $\bar{v}_\ga$.  The
function $\cU$ contains $2p+2$ oscillators $\theta$ which add up
to $d-2p-1$ oscillators $\theta$ in $(\Eop)^{d-2p-2}\chi$ and
generate the $\gep$-tensor by (\ref{LC_sym}). The operators
$\Eop$, $\chi$ and $\bta$ realize contractions of the  $\gep$-tensor
with the  background frame field, compensator and HS
curvatures, respectively. The operator $\bv$ realizes contractions
between the compensator and HS curvatures. The operator $\bs$
realizes contractions between two HS curvatures. Using the
symmetry of (\ref{action}) with respect to exchange of the $(p+1)$-form HS curvatures
we require $\cU(\bs,\bta,\bv)=\cU(\bs,\bta_i,\bta^i,\bv_i,\bv^i)$  to satisfy
the symmetry property
\be\label{u_symmetry}
\cU(\bs,\bta_i,\bta^i,\bv_i,\bv^i)=(-1)^{p+N+1}\cU(-\bs,\bta^i,\bta_i,\bv^i,\bv_i)\,,
\ee
where $N$ is defined by $\hat{\Go}_{(p)}(\psi)=(-1)^N\hat{\Go}_{(p)}(-\psi)$.

Taking into account (\ref{curv}) and (\ref{gauge+bianchi}), one can obtain (more details will be given in \cite{ASV_})
\begin{align}
\label{var}
\gd\cS_2&=2(-1)^{d+p}\int_{\cM^d}\lvac\diff\Big((\wedge\Eop)^{d-2p- 2}\chi\cU\Big)\wedge\hat{R}_{(p+1)}\wedge \delta\breve{\Omega}_{(p)}\rvac=\nn
\\
&=2(-1)^p\frac{\gl}{d-2p-1}\int_{\cM^d}\lvac(\wedge\Eop)^{d-2p-1}\chi\cQ\cU \wedge\hat{R}_{(p+1)}\wedge \delta\breve{\Omega}_{(p)}\rvac  \,,
\end{align}
where \be \label{Q}
\cQ=\Big(d-1+\bv^\ga\ptl{\bv^\ga}-\bta^\ga\ptl{\bta^\ga}\Big)\bv^\gb\ptl{\bta^\gb}+
\bs^{\ga\gb}\pptl{\bv^\ga}{\bta^\gb}\,. \ee The important fact is
that
\be
\label{Q_nilpotent}
\cQ^2=0\;,
\ee
which is a  consequence of $\diff^2=0$. A natural guess is that,
in an appropriate representation, $\cQ$ can be rewritten as a de Rham operator.
Indeed, one can see that
\be
\label{deRahm}
\delta =\bv^\ga\ptl{\bta^\ga}=A^{-1}\cQ A\,,
\ee
where
\be
\label{A}
A=(d-1+\bv^\ga\ptl{\bv^\ga}-\bta^\ga\ptl{\bta^\ga})!!
\exp(\frac{1}{2}\bs^{\ga\gb}\pptl{\bv^\ga}{\bv^\gb})\,.
\ee

Since the variation of the action (\ref{action}) has the form (\ref{var}), total derivative terms
in the action  result from  a $\cQ$-closed function $\cU$
that by virtue of  the Poincare Lemma can be represented as
\be
\label{total}
\cU(\bs,\bta,\bv)=\cQ \cT(\bs,\bta,\bv)
\ee
for some $\cT(\bs,\bta,\bv)$.

To find $\cU$ up to $\cQ$-closed terms we impose the extra field
decoupling condition requiring that the variation of the action
with respect to extra fields $w_{(p)}$ must vanish identically.
This is equivalent to the condition that  the resulting action
does not contain higher derivatives. However, it turns out that
this condition alone does not fix the action uniquely. The point
is that it is not enough to require the $AdS$ action to be invariant under
$AdS$ gauge symmetries and to contain first order derivatives in order to
guarantee that it describes a correct number of degrees of freedom. Correct choice
is dictated by the structure of the kinetic terms in the action
which, for the metric-type field $\Phi(x)$, is given by
${\cS}_2^{flat} \sim \int \d\Phi\d\Phi$ resulting from the $AdS_d$
action ${\cS}_2^{AdS} \sim \int \cD\Phi\cD\Phi + \lambda^2\Phi^2$
in the flat limit $\lambda \to 0$. In the $AdS_d$ space, mass-like terms
$\lambda^2\Phi^2$ break down all the gauge symmetries
(\ref{flat_g}) of the action ${\cS}_2^{flat}$ except for
$\delta\Phi = \cD S_p$ associated with the $AdS_d$ gauge parameter
$S_p$ (\ref{ads_g}). Thus an additional condition is that the
correct action must acquire all flat gauge symmetries in the flat
limit. As we show this is achieved if the action is independent of
the irrelevant auxiliary fields (more details will be given in
\cite{ASV_}). So we impose the decoupling condition requiring
 the variation with respect to extra fields $w_{(p)}$ and
irrelevant auxiliary fields $\go'_{(p)}$ to be identically zero
\be
\label{decouple}
\frac{\delta \cS_2}{\delta w_{(p)}} \equiv 0\,,
\qquad
\frac{\delta \cS_2}{\delta \go'_{(p)}} \equiv 0\,.
\ee
In other words the condition (\ref{decouple}) implies that the action
depends non-trivially on the physical field and the relevant
auxiliary field only. As we show in the rest of this letter the
decoupling condition fixes a form of the action up to a
normalization factor. Note that for symmetric HS fields \cite{LV}
and for particular examples of mixed-symmetry fields considered in
\cite{ASV,A2} irrelevant auxiliary fields are absent so that the
decoupling condition is equivalent to the extra field decoupling
condition \cite{LV,vf}.

To find the function $\cU$ we adhere
the following strategy. Firstly, we find the equations of motion in the form
consistent with the decoupling condition. Secondly, we reconstruct the action function $\cU$
that leads to these equations of motion.

For the subsequent analysis it is convenient to introduce a notion of weak equality.
Two polynomials $\cA(\bs,\bta,\bv)$ and $\cB(\bs,\bta,\bv)$
are  weakly equivalent  $\cA\sim \cB$, if
\be
\label{weak}
\lvac(\wedge\Eop)^{d-m-n}\chi\Big(\cA-\cB\Big)\wedge\hat{A}_{(m)}\wedge \breve{B}_{(n)}\rvac =0\,,
\ee
for any fields $A_{(m)}$ and $B_{(n)}$, which satisfy the Young symmetry and tracelessness  properties
(\ref{irr1})-(\ref{irr3}). The meaning of the weak equivalence of two functions is that they
coincide modulo terms proportional to Young symmetrizers and trace operators
which are zero by (\ref{irr1})-(\ref{irr3}).
A generic weakly zero function has the form
\begin{align}
\cW=&\sum_{i,j=1}^{s-1}\cW^{ij}\bs^{ij}+\sum_{i,j=1}^{s-1}\cW_{ij}\bs_{ij}+
\sum_{i,j=\!1,\,j>i}^{s-1}[\cW_i{}^j,l_i{}^j]+\sum_{i,j=\!1,\,j>i}^{s-1}[\cW^i{}_j,l^i{}_j]+\nn\\
\label{weakly_zero}{}+&\sum_{i=1}^{s-1}\Big([\cW_i,l_i{}^i]-\tilde{h}_i\cW_i\Big)+
\sum_{i=1}^{s-1}\Big([\cW^i,l^i{}_i]-\tilde{h}_i\cW^i\Big)\,,
\end{align}
where $\cW_{ij}$, $\cW^{ij}$, $\cW_i{}^j$, $\cW^i{}_j$, $\cW_i$ and
$\cW^i$ are arbitrary functions of $\bs$, $\bta$ and  $\bv$. Here the first two terms are weakly
zero due to the tracelessness condition (\ref{irr1}) and the other terms are weakly zero
due to the Young symmetry properties (\ref{irr2}), (\ref{irr3}). The important property is that the operators $\bs_{ij}$,
$\bs^{ij}$, $l_i{}^j$ and $l^i{}_j$ commute with $\cQ$. As a result,
$
\cQ\cW\sim 0\;, \forall\; \cW \sim 0\;.
$

\section{Equations of motion\label{Sct.:5}}

Let the variation (\ref{var}) have the form \be \label{eq} \delta{\cal
S}_2
=\int_{\cM^d}\lvac(\wedge\Eop)^{d-2p-1}\chi\cE(\bs,\bta,\bv)\wedge\hat{R}_{(p+1)}\wedge
\delta\breve{\Omega}_{(p)}\rvac \,, \ee
for some polynomial $\cE(\bs,\bta,\bv)$.
For the function $\cE$ to result from the variation of some
action (\ref{action}), it must be weakly $\cQ$-exact
\be\label{Qexactness_}
\cE\sim\cQ\cU\,.
\ee
Since
$\cQ^2=0$ and $\cQ$ maps weakly zero functions to weakly zero
functions, $\cE$ has to satisfy consistency condition
\be \label{Qcloseness} \cQ \cE \sim 0\,. \ee
Now our goal is to find $\cE$ satisfying the decoupling condition
(\ref{decouple}) and the consistency condition (\ref{Qcloseness}) and then to reconstruct $\cU$ via (\ref{Qexactness_}).

A function $\cE$ satisfying the decoupling condition has the form
\be
\label{funct_E}
\cE(\bs,\bta,\bv) = \Big(\bta_1\frac{\d}{\d \bv_1}
-\bta^1\frac{\d}{\d \bv^1}\Big)\tilde{\cE}(\bs,\bta)\bv^{2(s-1)}\;,
\ee
where $\bv^{2(s-1)}=\bv_1\cdots\bv_{s-1}\bv^1\cdots\bv^{s-1} $.
Actually, the first term in (\ref{funct_E})
contains $(s-1)$ compensators (which is a maximal possible number) contracted
with $\delta\Omega_{(p)}$
and therefore corresponds to the variation with respect to the physical field (cf. (\ref{phys})).
Analogously, the second term contains $(s-2)$ compensators contracted with
all columns of $\delta\Omega_{(p)}$ except for the first one and therefore
corresponds to the variation with respect to the relevant auxiliary field
(cf. (\ref{true_auxiliary})). In the both cases,
the remaining index in the first column of either $R_{(p+1)}$ or $\delta\Omega_{(p)}$
is contracted with the $\gep$-tensor by $\bta_1$ or $\bta^1$.
The relative coefficient in (\ref{funct_E})
is fixed by the symmetry property of  $\cU$ (\ref{u_symmetry}).

The important fact is that, using the Young symmetry properties
of the gauge fields, \textit{i.e.} by adding weakly zero terms, the function $\tilde{\cE}$ can always be chosen in the form
\begin{align}
\label{ee}
\tilde{\cE}(\bu,\bn)&=\Big(\prod_{i=1}^{s-1} (\bu_i)^{\tilde{h}_i-1}\Big)\!\!\!
\sum_{\stackrel{\scriptstyle
p_I\geq 0,\; I=1\div n} {\stackrel{\scriptstyle p_1+\cdots +p_n=p}{}}}\!\!\!
\rho(p_1,\ldots,p_n)\Big(\frac{\bn_{\mu_1}}{\bu_{\mu_1}}\Big)^{p_1}\cdots
\Big(\frac{\bn_n}{\bu_{\mu_n}}\Big)^{p_n}\,,
\end{align}
where $\mu_I$ is a number of the first column of the $I$-the block ({\it i.e.} $\mu_1=1$, $\mu_2=m_1+1$,..., $\mu_n=m_1+\cdots+m_{n-1}+1$) and new variables
\begin{align}
\bu_i&=\bs_i{}^i\,,\qquad \bn_I=\bta_{\mu_I}\bta^{\mu_I}\,,
\end{align}
(no sums over repeated indices) realize column-to-column contractions between $R_{(p+1)}$ and $\gd\Go_{(p)}$ and contractions
of the $\gep$-tensor with the first columns of $I$-th vertical blocks of
$R_{(p+1)}$ and $\gd\Go_{(p)}$. The coefficients $\rho(p_1,\ldots,p_n)$ parameterize
various types of contractions between $2p$ indices of the $\gep$-tensor and those of $R_{(p+1)}$,
$\gd\Go_{(p)}$.

Let us now solve the equation (\ref{Qcloseness}). Leaving details for
\cite{ASV_} we give the final result of the
substitution of the ansatz (\ref{funct_E}) into the equation (\ref{Qcloseness}). Modulo weakly zero terms we obtain the
following system of equations
\begin{align}
\label{system}
&\left( \frac{(\bar{N}_1+2)(\bar{N}_I+m_I)}{\bar{U}_1(\bar{N}_I+1)}
\bu_{\mu_1}\ptl{\bn_1}
+\sum_{J=2}^{I-1}\;\frac{(\bar{N}_J+1)(\bar{N}_I+m_I)}
{\bar{U}_J(\bar{N}_I+1)}\bu_{\mu_J}\ptl{\bn_J}+\nn\right.\\
&\left.+\Big(\sum_{J=I+1}^{n}\frac{\bar{N}_J}{\bar{U}_I}
-\frac{s-\mu_I-m_I}{\bar{U}_I}-1\Big)\bu_{\mu_I}\ptl{\bn_I}\right) \tilde{\cE}(\bu,\bn)=0\,,
\qquad 2\leq I\leq n\,,
\end{align}
where
\begin{align}
\bar{N}_I=\bn_I\ptl{\bn_I}\,,\qquad \bar{U}_I=\bu_{\mu_I}\ptl{\bu_{\mu_I}}\;,\quad
\hbox{no summation}\,.
\end{align}
Taking into account (\ref{ee}), eq. (\ref{system}) can be
reduced to a system of recurrent equations
\be
\label{recurrent1}
\rho(p_1-1, ... ,p_I+1,..., p_n)= G_I(p_1,\ ...,
p_I,...,p_n)\,\rho(p_1,...,p_I,..., p_n)\,,\;\; I=2\div
n\,,
\ee
\be
\label{range}
p_1\geq 1\;,\qquad p_I\geq 0\;,\;\; I=2\div n\;,\qquad p_1+ ... + p_n=p\;,
\ee
where
\begin{align}
\label{Green} G_I(p_1,..., p_n)=
\frac{(p_I-h_I+1)(p_I+m_I)}{(p_I+1)^2}\frac{p_1(p_1+1)}{h_1-p_1}
\frac{\prod\limits_{J=2}^{I-1}\Big(\sum\limits_{K=J+1}^{n}p_K-\vartheta(J)-m_J\Big)}
{\prod\limits_{J=2}^{I}\Big(\sum\limits_{K=J}^{n}p_K -
\vartheta(J)\Big)}\,
\end{align}
and $\vartheta(I) = s-\mu_I-m_I+h_I-1$. Note that the arguments of the factorials in the numerators
of (\ref{Green}) are strictly non-negative in the range (\ref{range}). The general solution is
\be
\label{gensol}
\ba{l}
\rho(p_1,..., p_n) =
\\
\dps
=\frac{\rho\,\delta(p-\sum\limits_{I=1}^{n}p_I)}{(p_1!)^2(p_1+1)(h_1-p_1-1)!}\prod_{I=2}^{n}\frac{(p_I+m_I-1)!}{(p_I!)^2\,(h_I-p_I-1)!}
\frac{\Big(\vartheta(I)-\sum\limits_{J=I}^{n}p_J\Big)!}
{\Big(\vartheta(I)+m_I-\sum\limits_{J=I+1}^{n}p_J\Big)!}\,,
\ea
\ee
where  $\rho$ is an arbitrary constant. It is elementary to check that the function
(\ref{gensol}) does solve the system (\ref{recurrent1})-(\ref{Green}).

Expressions  (\ref{funct_E}), (\ref{ee}), (\ref{gensol}) determine function $\cE$
satisfying the weak $\cQ$-closedness equation (\ref{Qcloseness}) and give rise to the variation
(\ref{eq}) satisfying the decoupling condition (\ref{decouple}).
The constructed $\cE(\bs,\bta,\bv)$ is (weakly) unique up to a normalization factor $\rho$.

\section{Reconstruction of the action \label{Sct.:6}}

Having found the function $\cE$ we now solve the eq. (\ref{Qexactness_}) on
the action function $\cU$.  To this end, we rewrite the  $\cQ$-closedness
condition (\ref{Qcloseness}) as a strong equality
\be
\label{Qcloseness_}
\cQ\cE=\cP\,,
\ee
where $\cP$ represents some weakly zero  terms
\be\label{Pwz}
\cP\sim 0\,.
\ee
From (\ref{Qcloseness_}) we have
that $\cQ\cP=0$.
As the operator $\cQ$ (\ref{deRahm}) is equivalent to the de Rham operator $\gd$, by Poincare Lemma  the
function $\cP$ is $\cQ$-exact
\be\label{K}
\cP=\cQ\cK\,.
\ee

Let us show that $\cK$ in (\ref{K}) can be chosen to be weakly zero.
Indeed, taking into account (\ref{deRahm}), we rewrite eq. (\ref{K}) as
\be\label{K_}
\cP'=\gd\cK'\,,
\ee
where $\cP'=A^{-1}\cP$, $\cK'=A^{-1}\cK$ and the operator $A$ is given by (\ref{A}).
Consider the operator $\gd^*$
\be
\gd^*=\bta^\ga\ptl{\bv^\ga}\;, \qquad \gd^{*\,2}=0\,.
\ee
One obtains that
\be
\Gd\equiv \{\gd,\gd^*\}=\bta^\ga\ptl{\bta^\ga}+\bv^\ga\ptl{\bv^\ga}\,.
\ee
Acting by $\gd^*$ at the both sides of (\ref{K_}) one obtains
\be
\Gd\cK'=\gd^*\cP'+\gd\gd^*\cK'\,.
\ee
The operator $\Gd$ commutes with $\gd$ and $\gd^*$. As a result, a  partial solution of (\ref{K_}) is
\be\label{K__}
\cK'=\Gd^{-1}\gd^*\cP'\,.
\ee
The operator $\Gd^{-1}$ admits the following integral realization
\be\label{Delta-1}
\Gd^{-1}\cA(\bs,\bta,\bv)=\int_0^1 \!\frac{{\rm d}t}{t}\cA(\bs,t\bta,t\bv)\,
\ee
for a function $\cA(\bs,\bta,\bv)$ such that $t^{-1}A(\bs, t\bta,t\bv)$ is polynomial in $t$.
Substituting (\ref{Delta-1}) into (\ref{K__}) one obtains
\be
\cK'(\bs,\bta,\bv)=\int_0^1 \!\frac{{\rm d}t}{t}\bta^\gb\ptl{\bv^\gb}
\cP'(\bs,t\bta,t\bv)\,.
\ee
Equivalently,
\be
\label{K___}
\cK(\bs,\bta,\bv)=A\Big\{\int_0^1 \!\frac{{\rm d}t}{t}\bta^\gb\ptl{\bv^\gb}
\Big(A^{-1}\cP\Big)(\bs,t\bta,t\bv)\Big\}\,.
\ee
As the operators $\gd^*$, $A$, $A^{-1}$, $\Gd$, $\Gd^{-1}$ commute
with the operators $\bs_{ij}$, $\bs^{ij}$, $l_i{}^j$, $l^i{}_j$ and $\cP$ is weakly zero (\ref{Pwz}),
we conclude from (\ref{weakly_zero}) that $\cK$ of the form (\ref{K___}) is weakly zero as well.

Now we are in a position to solve eq. (\ref{Qexactness_}). Rewriting eq.
(\ref{Qexactness_}) as a strong equality
\be
\label{Qcloseness__}
\cQ\cU=\cE - \cK\,,
\ee
we find analogously to (\ref{K_}) that
\be\label{U__}
\cU(\bs,\bta,\bv)=A\Big\{\int_0^1 \!\frac{{\rm d}t}{t}\bta^\gb\ptl{\bv^\gb}
\Big(\Big(A^{-1}\cE\Big)(\bs,t\bta,t\bv)-\Big(A^{-1}\cK\Big)(\bs,t\bta,t\bv)\Big)\Big\}\,.
\ee
Substituting (\ref{U__}) into the action (\ref{action}) one finally obtains
\be
\label{action__}
\cS_2=\gl^{-2(s-1)}\int_{\cM^d}\lvac(\wedge\Eop)^{d-2p-2}\chi A\Big\{\int_0^1 \!\frac{{\rm d}t}{t}\bta^\gb\ptl{\bv^\gb}
\Big(A^{-1}\cE\Big)(\bs,t\bta,t\bv)\Big\}\wedge\hat{R}_{(p+1)}\wedge \breve{R}_{(p+1)}\rvac\,,
\ee
where the weakly zero term with $\cK$ does not contribute to the action and the
factor of $\gl^{-2(s-1)}$ is introduced to provide correct flat limit.
The action (\ref{action__}) satisfies the decoupling condition and is defined uniquely modulo total derivatives (\ref{total}).

\section{Flat space gauge symmetries\label{Sct.:7}}

As explained in sect. 4  the constructed
$\ads$ HS action (\ref{action__}) should describe correct
dynamics of a mixed-symmetry gauge field in the flat limit $\lambda=0$
and hence has to be invariant under additional set of gauge symmetries
(\ref{flat_g}). Here we show that in the flat limit   of the $\ads$ action (\ref{action__})
a gauge symmetry enhancement indeed takes place with respect to traceless flat space gauge parameters.

Using the standard gauge $V^A=\delta^A{}_{d+1}$ \cite{compensator},
setting the frame field to $E_{0\un}^a=\gd_\un^a$, Lorentz spin connection to $\omega^{ab}_{\un}=0$ and
replacing the Lorentz covariant derivative $\cD$ with the flat derivative $\d$, one rewrites
the variation (\ref{eq}) in the form
\begin{align}
\label{flat_variation}
\gd\cS_2^{\rm flat}=\int_{\cM^d}\Big(&\lvac(\wedge\Eop)^{d-2p-1}\,\chi \tilde{\cE}(\bu,\bta)\,
\bta^1\,\wedge \hat{r}_{(p+1)}\wedge \delta\breve{\omega}_{(p)}\rvac-\nn\\
{}-&\lvac(\wedge\Eop)^{d-2p-1}\,\chi \tilde{\cE}(\bu,\bta)\,\bta_1\,\wedge\hat{\cR}_{(p+1)}\wedge
\delta\breve{e}_{(p)}\rvac\Big)\,,
\end{align}
where the $(p+1)$-forms
$r_{(p+1)}=\extdiff e_{(p)}+\cdots$ and $\cR_{(p+1)}=\extdiff \go_{(p)}+\cdots$
are the Lorentz components of the curvature $R_{(p+1)}$ associated with the
physical and relevant auxiliary fields, respectively.
The variation over the relevant auxiliary field $\go_{(p)}$ gives rise to the equation  of motion
which can be cast into the following component form (for more details see \cite{ASV_})
\be\label{equation_w}
r^{a_1[\tilde{h}_1-1],\ldots,a_{s-1}[\tilde{h}_{s-1}-1];\,m[p+1]}(x)=C^{a_1[\tilde{h}_1-1],
\ldots,a_{s-1}[\tilde{h}_{s-1}-1], m[p+1]}(x)\,,
\ee
where world indices are converted into the tangent indices $m$.
It was argued in Ref. \cite{ASV} that the
tensor $C^{a_1[\tilde{h}_1-1],\ldots,a_{s-1}[\tilde{h}_{s-1}-1], m[p+1]}(x)$ is either zero
if $\tilde{h}_{s-1}=p+1$ or equals to the primary Weyl tensor if $\tilde{h}_{s-1}\neq p+1$. Equation (\ref{equation_w}) expresses the relevant auxiliary field
$\go_{(p)}$ in terms of first derivatives of the physical field $e_{(p)}$  up to pure gauge
degrees of freedom. It is convenient to use the $1.5$-formalism with the auxiliary field expressed implicitly
by virtue of its equation of motion through the physical field.

The next step is to study how the frame-type physical field is transformed under the
flat gauge symmetries. The flat gauge parameters $S_{(I)}, I=1\div n_\Phi$  (\ref{flat_g}) result from cutting a cell from
$I$-th block of the metric-type field $\Phi(x)$. As a result, there is as many independent flat gauge parameters as the
number of different vertical blocks of the Young tableau corresponding to the field $\Phi(x)$.
Note that the gauge parameter $S_{(n_\Phi)}$ associated with the vertical
block of the minimal height corresponds to the physical Lorentz-covariant component of the $\ads$
gauge parameter $\xi_{(p-1)}$ (\ref{adsgaugetr}).

It is convenient to convert all world indices  originally carried by differential forms into
tangent ones, contracting them in the operators $\hat{A}_{(n)}$=($\delta\hat{\Omega}_{(p)}$, $\hat{R}_{(p+1)}$, $\hat{S}_{(I)}$) and
$\breve{A}_{(n)}$=($\delta\breve{\Omega}_{(p)}$, $\breve{R}_{(p+1)}$, $\breve{S}_{(I)}$)
with the oscillators $\kappa^\bullet_a$, $\bkappa_\bullet^a$ and $\kappa_\bullet^a$, $\bkappa^\bullet_a$,
which anticommute  with the previously introduced oscillators
and satisfy the anticommutation relations
\be
\label{kappa}
\{\bkappa_\bullet^a,\kappa^{\bullet b}\}=\eta^{ab}\,,\qquad
\{\bkappa^{\bullet a},\kappa_\bullet^ b\}=\eta^{ab}\,,
\ee
with other anticommutators being zero.
The Fock vacua are defined by
$\lvac\kappa_\bullet^a=0$, $\lvac\kappa^{\bullet a}=0$, $\bkappa_{\bullet}^a\rvac=0$ and
$\bkappa^{\bullet a}\rvac=0$.

A flat gauge transformation of the metric-type field $\Phi(x)$ results as an appropriate projection
of the following transformation of the physical field $e_{(p)}$
\begin{align}
\label{di_e}
\gd_{(I)}\breve{e}_{(p)}\rvac &=\cP D^{\mu_I}\breve{S}_{(I)}\rvac\,,
\end{align}
where $D_{\mu_I} = \psi_{\mu_I}^a\d_a$ and  $\cP$ projects r.h.s. of (\ref{di_e}) on the Young tableau
associated with tangent indices of the field $e_{(p)}$.

Substituting (\ref{di_e}) into (\ref{flat_variation}) and neglecting terms with the variation of the auxiliary
field by using the 1.5-order formalism, one gets the variation with respect to
the parameter $S_{(I)}$
\be
\label{variation_exi}
\gd_{(I)}\cS_2^{\rm flat}=\int_{\cM^d}\Big(\ga_1\Delta_1+\ga_2\Delta_2+\ga_3\Delta_3\Big)\,,
\ee
where $\ga_{1,2,3}$ are some coefficients and
\begin{align}
\Delta_1&=\frac{1}{p}\lvac(\Eop)^d\chi \cF_{(I)}(\bu)\bar{D}_1\bar{D}_I(\bar{K})^p\hat{\omega}_{(p)}
\breve{S}_{(I)}\rvac\,,\nn\\ \nn\\
\Delta_2&=-\lvac(\Eop)^d\chi \cF_{(I)}(\bu)\bar{D}_1\bar{M}_I\tilde{D}(\bar{K})^{p-1}
\hat{\omega}_{(p)}\breve{S}_{(I)}\rvac\,,\\ \nn\\
\Delta_3&=\lvac(\Eop)^d\chi \cF_{(I)}(\bu)\bar{M}_1\bar{D}_I\tilde{D}(\bar{K})^{p-1}
\hat{\omega}_{(p)} \breve{S}_{(I)}\rvac\,\nn
\end{align}
with
\be
\bar{D}_I =\bpsi_{\mu_I}{}^a\d_a\,,
\;\; \tilde{D}= \bar{\kappa}^\bullet_a\d^a\,,
\;\; \bar{M}_I=\bkappa_\bullet^a\bpsi_{\mu_Ia}\,,
\;\; \bar{K} = \bkappa_\bullet^a\bkappa^\bullet_a\,,
\;\; \cF_{(I)}(\bu) = \frac{1}{\bu_{\mu_I}}\prod_{i=1}^{s-1}(\bu_i)^{\tilde{h}_i-1}.
\ee
Then one can show  that the coefficients $\alpha_{1,2,3}$ satisfy the  linear relation
\be
\label{alpha_rel}
\ga_1+\ga_2+\ga_3=0
\ee
resulting from the invariance of the original $AdS_d$ action under Lorentz-type  gauge symmetry acting on the frame-type
HS field with the Lorentz gauge parameter $\xi_{(p-1)}$  of the same Young symmetry type as the
relevant auxiliary field.

On the other hand, the operators $\Delta_{1,2,3}$ satisfy relations resulting from Bianchi identities.
Indeed, consider the particular Bianchi identity at $\lambda=0$
\be
\label{bianchi}
\extdiff r_{(p+1)} +\sigma_{-}\cR_{(p+1)}+... =0\;.
\ee
Here $\sigma_{-}$ is the operator decreasing a number of Lorentz indices and dots stand for
contributions of the curvatures of the same Young symmetry as irrelevant auxiliary fields
that can be disregarded in the flat limit by virtue of the decoupling condition (\ref{decouple}).
Taking into account the eq. (\ref{equation_w})
and projecting out the contribution of the Weyl tensor into Bianchi identity
(\ref{bianchi}) one can prove that
\be
\label{finish}
\Delta_1=\Delta_2=\Delta_3
\ee
(more details will be given in \cite{ASV_}).
Comparing the relations (\ref{alpha_rel}) and (\ref{finish}) one concludes that the variation (\ref{variation_exi}) is zero.
Thus the constructed $\ads$ action functional (\ref{action__})
exhibits in the flat limit additional gauge symmetries associated with the corresponding
mixed-symmetry gauge field on Minkowski space.

\section{Conclusions\label{Sct.:8}}

In this letter we have announced the covariant Lagrangian formulation for a
generic mixed-symmetry bosonic gauge field propagating on the
$\ads$ background and corresponding to a unitary $o(d-1,2)$-module.
A novel feature of  gauge fields of mixed-symmetry type is that, apart from the previously known extra
field decoupling condition, it is necessary to require all but one auxiliary fields also not to contribute
to the action. The relevant auxiliary field contributing to the action is the maximally
antisymmetric one. This condition fixes the Lagrangian  modulo total
derivative terms and guarantees correct flat gauge symmetry enhancement that, in its turn,
implies correct counting of degrees of freedom.

The next step for the further study is to generalize the results of this letter
to the case of fermionic fields. Also the unfolded formulation for mixed-symmetry fields is to be developed
since it clarifies a structure of  non-Abelian HS symmetries
and of a non-linear HS theory with mixed-symmetry HS gauge fields.
We plan to discuss these problems in the future publication \cite{ASV_}.

\vspace{2mm}

{\bf Acknowledgements.} This work is supported by grants RFBR No 02-02-17067, LSS No
1578-2003-2, INTAS No 03-51-6346.

\end{document}